\documentclass[11pt,a4paper]{article}
\usepackage{jheppub}
\usepackage{amsmath,amsfonts,amsthm,graphicx,ulem}
\usepackage{float}
\newcommand{\bal}{\begin{align}}
\newcommand{\eal}{\end{align}}
\newcommand{\beq}{\begin{equation}}
\newcommand{\eeq}{\end{equation}}
\newcommand\beqa{\begin{eqnarray}}
\newcommand\eeqa{\end{eqnarray}}
\newcommand\bea{\begin{array}}
\newcommand\eea{\end{array}}

\renewcommand{\leq}{\leqslant}
\renewcommand{\geq}{\geqslant}

    \newcommand{\COMMENT}[1]{}
    
    \newcommand{\neqa}{\nonumber\end{eqnarray}}

\def\[{\left[}
\def\]{\right]}

\def\D{\Delta}

\def\\langle{\langle}
\def\>{\rangle}

\def\i2{\frac{i}{2}}

\newcommand{\BC}{$\textit{BC}_2$}

\newcommand{\dcoset}{\mathbin{\!/\mkern-5mu/\!}}
\renewcommand{\L}{\mathcal{L}}
\newcommand{\R}{\mathcal{R}}

\title{Harmony of Spinning Conformal Blocks}
\author[a]{Volker Schomerus,}
\author[b]{Evgeny Sobko,}
\author[c]{Mikhail Isachenkov}

\affiliation[a]{DESY Hamburg, Theory Group, Notkestra{\ss}e 85, 22607 Hamburg, Germany}
\affiliation[b]{Nordita and Stockholm University,
Roslagstullsbacken 23, SE-106 91 Stockholm, Sweden}
\affiliation[c]{Department of Particle Physics and Astrophysics, Weizmann Institute of Science, Rehovot 7610001, Israel}
\emailAdd{evgenysobko AT gmail.com}
\emailAdd{volker.schomerus AT desy.de}
\emailAdd{m.isachenkov AT gmail.com}

\abstract{Conformal blocks for correlation functions of tensor operators play an increasingly important role for the conformal bootstrap programme. We develop a universal approach to
such spinning blocks through the harmonic analysis of certain bundles over a coset of the
conformal group. The resulting Casimir equations are given by a matrix version of the
Calogero-Sutherland Hamiltonian that describes the scattering of interacting spinning particles in a 1-dimensional external potential. The approach is illustrated in several examples including fermionic seed blocks in 3D CFT where they take a very simple form.}

\subheader{DESY-16-239, NORDITA-2016-123,WIS/07/16-DEC-DPPA}

\usepackage{varioref}
\usepackage{makeidx}
\makeindex
\usepackage{amssymb}

\begin{document}

  \maketitle
\def\SO{\textrm{SO}}

\section{Introduction}

The conformal bootstrap programme, which was originally formulated in the \cite{Ferrara:1973yt,Polyakov:1974gs,Mack:1975jr}, has raised hopes for a
new non-perturbative construction of conformal field theories in any
dimension, even of theories for which an action cannot be written down or
a microscopic (UV) description is not known. The programme rests on a
careful separation of kinematical and dynamical data in correlation
functions, i.e. on the split into the kinematical conformal
blocks and the dynamical coefficients of the operator product
expansion. The latter are severely constrained by the so-called
crossing symmetry equations, an infinite set of coupled equations
for the operator product coefficients with kinematically determined
coefficients. Over the last few years, numerical studies of these
crossing symmetry equations have given access to critical exponents
and operator product coefficients with enormous precision
\cite{Rattazzi:2008pe, ElShowk:2012ht, El-Showk:2014dwa,
Simmons-Duffin:2015qma, Kos:2016ysd}.

While initial work has focused on correlation functions involving one or
two scalars, tensor fields are only now beginning to receive some attention
in the bootstrap programme. The most important tensor field is clearly the stress
tensor which, by definition, exists in any conformal field theory. If the
conformal blocks for tensor fields were under good control one could
explore the space of conformal field theories without assumptions on the
scalar subsector. The study of such spinning conformal blocks was initiated
in \cite{Costa:2011mg,Costa:2011dw}. A fairly generic approach was proposed
in \cite{SimmonsDuffin:2012uy}, based on the so-called shadow formalism of
Ferrara et al. \cite{Ferrara:1972ay,Ferrara:1972uq,Ferrara:1972xe,
Ferrara:1973vz}, see also \cite{Costa:2016hju,Costa:2016xah} for more
recent work and further references. This leads to expressions in which
conformal blocks are simply sewn together from 3-point functions. In
the bootstrap programme, such formulas are difficult to work with, partly
because they involve a the large number of integrations. On the other
hand, recent work \cite{Echeverri:2016dun} clearly shows that explicit
constructions of spinning blocks in higher dimensional conformal field
theories in terms of known special functions are possible. The main
motivation for our work is to pave the way for systematic extensions
of such efficient formulas.

In order to achieve this, we generalize an interesting interpretation of
conformal blocks as wave functions of an interacting 2-particle Schr\"odinger
problem with Calogero-Sutherland potential that was recently uncovered
in \cite{Isachenkov:2016gim}. More precisely, it was shown that the Casimir
equations for scalar conformal blocks \cite{Dolan:2003hv} are equivalent to
the eigenvalue equations for a hyperbolic Calogero-Sutherland Hamiltonian.
The integrability of this Hamiltonian has been argued to provide a new avenue
to scalar conformal blocks. Only very few Casimir equations for spinning
blocks have been worked out in the literature, see however
\cite{Iliesiu:2015akf,Echeverri:2016dun}. Here we propose an independent
approach that allows us to construct an appropriate Calogero-Sutherland
model for any choice of external
operators with spin. In comparison to the case of scalar blocks, the potentials
become matrix valued and describe the motion of two interacting particles
with spin in a 1-dimensional (spin-dependent) external potential. The
associated Schr\"odinger problems are equivalent to the Casimir equations
for spinning blocks.

Let us describe the main results and plan of this paper. Throughout the next two
sections we shall set up a model for spinning conformal blocks in any dimension
where the 4-point blocks are represented as sections in a certain vector bundle
over the following double coset of the conformal group \(G=\SO(1,d+1)\)
$$ \mathcal{C}= \SO(1,d+1)\dcoset (\SO(1,1)\times \SO(d))\ .   $$
The denominator consists of dilations and rotations and we divide by both its right
and its left action on the conformal group. As we shall argue in section 4, this
coset space is 2-dimensional and parameterizes the conformally invariant cross
ratios. Let us notice that, once we have divided by the right action, the left
action of $\SO(1,1) \times \SO(d)$ in the quotient is stabilized by a subgroup
$\SO(d-2) \subset \SO(d)$ of the rotation group.

Given four tensor fields that transform in representations with highest weight $\mu_i,
i = 1, \dots, 4$ of the rotation group $\SO(d)$, the fibers of the relevant bundles
over the double coset $\mathcal{C}$ are given by
$$ T =  \left(V_{\mu_1} \otimes V_{\mu'_2} \otimes
       V_{\mu_3} \otimes V_{\mu'_4}\right)^{\SO(d-2)}\ . $$
Here, $V_{\mu}$ denotes the carrier space of the representation $\mu$ of the rotation
group.\footnote{Strikly speaking, the fibers also carry an action of dilations that is
determined by the values of the external conformal weights. We will specify this later.}
Also we used the prime symbol for the representation
\(\mu'(r)=\mu(wrw)\) that is twisted by conjugation with the nontrivial element \(w\) of restricted Weyl
group. The latter is given by the quotient \(R'/R\) where \(R'\) is a normaliser of the dilation subgroup
\(D\) within the maximal compact subgroup \(\SO(d+1)\) and it consists of two elements $\{1,w\}$. Representation space \(V_{\mu'}\) coincides with \(V_\mu\) and we add prime symbol in order to stress that we consider this vector space as carrier of the representation \(\mu'\). We consider the tensor product inside the brackets as a representation of the
subgroup $\SO(d-2)\subset \SO(d)$ and select the subspace of $\SO(d-2)$ invariants.
As we shall argue in section 2 and 3, elements of the resulting vector space should be
considered as 4-point tensor structures. We stress that the global structure of the
relevant vector bundles also depends on the choice of conformal weights. As described
in section 3, the bundle can be realised as a space of equivariant functions over
\(G\) which defined by their restriction to \(\mathcal{C}\).

Once the model of conformal blocks is set up, we derive the relevant Casimir equations
for spinning blocks from the Laplacian on the conformal group $\SO(1,d+1)$ in section 4.
Following the logic of Hamiltonian reduction described in \cite{Feher:2007ooa,Feher:2009wp}, we argue that these equations can be brought into a matrix
Schr\"odinger problem for two interacting particles with spin that are moving in a 1-dimensional external potential. In the case of non-trivial fermionic seed blocks in 3-dimensional conformal field theory, the relevant Hamiltonian is worked out explicitly, see section 5. It is associated with a 4-point correlation function of two scalars and two spin-$1/2$ fermions
\footnote{Strictly speaking, spin-$1/2$ fermions are in a representation of the universal
covering group Spin$(3)$ = SU$(2)$ of the rotation group $R =\, $SO$(3)$. Throughout most
of this text we shall not distinguish between Spin$(d)$ and SO$(d)$.},
i.e.\ two of the Spin$(3) = $ SU$(2)$ representations $\mu$ are 1-dimensional while
the other two are 2-dimensional. The fiber $T$ of our bundle is 4-dimensional and the Hamiltonian has block-diagonal form \(H=\text{diag}\{-2\mathcal{M}_{1},-2\mathcal{M}_{2}\}\) with the following entries
\begin{gather}\label{CasimirNewVar}
\mathcal{M}_{1}=\begin{pmatrix}
H_{CS}^{(a,b,1)}+\frac{5}{4} & 0 \\
0 & H_{CS}^{(a,b,1)}+\frac{5}{4}
\end{pmatrix}+\\
\begin{pmatrix}
  \frac{-1}{16}(\frac{1}{\cosh^2\frac{x}{2}}+\frac{1}{\cosh^2\frac{y}{2}}
  -\frac{2}{\sinh^2\frac{x-y}{4}}-\frac{2}{\sinh^2\frac{x+y}{4}}) &
  \frac{a+b}{4}(\frac{1}{\cosh^2\frac{x}{2}}-\frac{1}{\cosh^2\frac{y}{2}}) \\
  \frac{a+b}{4}(\frac{1}{\cosh^2\frac{x}{2}}-\frac{1}{\cosh^2\frac{y}{2}}) &
  \frac{-1}{16}(\frac{1}{\cosh^2\frac{x}{2}}+\frac{1}{\cosh^2\frac{y}{2}}
  +\frac{2}{\cosh^2\frac{x-y}{4}}+\frac{2}{\cosh^2\frac{x+y}{4}})
 \end{pmatrix}
 \notag \\
\mathcal{M}_{2}=\begin{pmatrix}
H_{CS}^{(a,b,1)}+\frac{5}{4} & 0 \\
0 & H_{CS}^{(a,b,1)}+\frac{5}{4}
\end{pmatrix}+\\
\begin{pmatrix}
  \frac{1}{16}(\frac{1}{\sinh^2\frac{x}{2}}+\frac{1}{\sinh^2\frac{y}{2}}
  +\frac{2}{\sinh^2\frac{x+y}{4}}-\frac{2}{\cosh^2\frac{x-y}{4}}) &
  \frac{b-a}{4}(\frac{1}{\sinh^2\frac{x}{2}}-\frac{1}{\sinh^2\frac{y}{2}}) \\
  \frac{b-a}{4}(\frac{1}{\sinh^2\frac{x}{2}}-\frac{1}{\sinh^2\frac{y}{2}}) &
  \frac{1}{16}(\frac{1}{\sinh^2\frac{x}{2}}+\frac{1}{\sinh^2\frac{y}{2}}
  +\frac{2}{\sinh^2\frac{x-y}{4}}-\frac{2}{\cosh^2\frac{x+y}{4}})
\end{pmatrix}\notag
\end{gather}
where \(H_{CS}^{(a,b,1)}\) is a Calogero-Sutherland Hamiltonian of \BC\ type,
see eq.\ \eqref{sCSHam}. In the appendix A we map this Hamiltonian to the set of
Casimir equations for 3D fermionic seed blocks that were worked out in
\cite{Iliesiu:2015akf}. Our matrix Hamiltonian describes the two spin-$1/2$
particles in a 1-dimensional external potential with an infinite wall at
$x=0,y=0$. The interaction of the particles with the wall depends on the
spin and it can induce spin flips, i.e. involves off-diagional terms, if
the parameters $a \neq 0$ or $b\neq 0$. In addition, the particles possess
a spin-dependent interaction. The latter is purely diagonal.

The paper finally concludes with a list of open questions and further directions.
Among them are the analysis of Casimir equations in dimension $d\geq 4$, the study
of boundary and defect blocks as well as spinning blocks for non-BPS operators in
superconformal field theories. Integrability and solutions of the Casimir equations
are briefly commented on while details are left for future research.

\section{Conformal blocks and Tensor Structures}\label{Sec.ConfBlocks}

In this section we shall review the basic model of spinning conformal blocks
in the context of 4-point correlation functions on $\mathbb{R}^d$. We will work
in Euclidian \(d\)-dimensional space so that the conformal group is \(G=\SO(1,d+1)\).
Primary fields of a conformal field theory sit in representations $\chi_\pi$ of $G$
that are induced from a representation $\pi$ of the subgroup
$K = \SO(1,1) \times \SO(d)\subset G$. Here, the factor $D=\SO(1,1)$ is generated by
dilations while $R=\SO(d)$ consists of all rotations $r$ of the d-dimensional
Euclidean plane. The choice of $\pi$ encodes the conformal weight $\Delta$ and the
highest weight $\mu$ of the rotation group \(\SO(d)\). We shall use $\pi =
\pi^\Delta_\mu$ to display the dependence on $\Delta$ and $\mu$. From time to time
we will also write $\pi = (\Delta,\mu)$.

It is well known that the correlation functions of two primary
operators are uniquely fixed (up to normalization) by conformal symmetry
to take the following form
\begin{gather}\label{2ptTS}
\langle\mathcal{O}_i(x_1)\mathcal{O}^\dag_j(x_2)\rangle=\frac{\delta_{ij}
t_{ij}}{|x_{12}|^{2\Delta_i}}
\end{gather}
where \(t\) is a unique tensor structure. As an example consider correlation
function of two primary operators \(\mathcal{O}^{\nu_1...\nu_l}\)  which
transforms as symmetric traceless tensors under the action of the rotation group
$R=\SO(d)$. It is customary to contract the indices of such fields with the
indices of a lightlike vector $\zeta_\nu$, i.e. to introduce
$$ \mathcal{O}(x;\zeta) \equiv \mathcal{O}^{\nu_1 \cdots \nu_l}(x) \,
\zeta_{\nu_1} \cdots \zeta_{\nu_l}
\ . $$
The corresponding 2-point functions can be written as
\begin{gather}
\langle\mathcal{O}(x_1,\zeta_1)\, \mathcal{O}(x_2,\zeta_2)^\dag\rangle
=\frac{1}{|x_{12}|^{2\Delta}}\left(\zeta_{1,\nu} I^{\nu \eta} \zeta_{2,\eta}\right)^l
\end{gather}
where \(I^{\nu \eta}=g^{\nu \eta}-2x^{\nu}_{12} x_{12}^{\eta}/|x_{12}|^2\). Correlation
function of three primary operators corresponding to representations $(\Delta_1,\mu_1),
(\Delta_2,\mu_2)$ and $(\Delta_3,\mu_3)$ can be written as a sum over conformally
invariant tensor structures $t^\alpha$
\begin{gather}\label{3point}
\langle\mathcal{O}_1(x_1)\mathcal{O}_2(x_2)\mathcal{O}_3(x_3)\rangle
= \frac{\sum_{\alpha=1}^{N_3}\lambda^\alpha_{123}t^\alpha_{123}}{|x_{12}|^{\Delta_{12,3}}
|x_{23}|^{\Delta_{23,1}}|x_{13}|^{\Delta_{13,2}}},
\end{gather}
where $\Delta_{12,3} = \Delta_1+\Delta_2-\Delta_3$ etc. and  \(N_3= N_3(\mu_1,\mu_2,\mu_3)\)
denotes the number of tensor structures \(t^\alpha\) that can appear. Finally,
\(\lambda^\alpha_{123}\) are the structure constants that are not determined by conformal
symmetry and carry dynamical information. Note that we have suppressed all tensor indices
in eq.\ \eqref{3point}. In case two of the fields, let's say $\mathcal{O}_1$ and $\mathcal{O}_2$
are scalar and the field $\mathcal{O}_3$ is a symmetric traceless tensor of spin $l$, there is a
unique tensor structure, i.e.\ $N_3 =1$, and the correlator reads
\begin{gather}
\langle\mathcal{O}_1(x_1)\mathcal{O}_2(x_2)\mathcal{O}_3(x_3,\zeta)\rangle
= \frac{\lambda^\alpha_{123} Z^l}{|x_{12}|^{\Delta_{12}}|x_{23}|^{\Delta_{23}}|x_{13}|^{\Delta_{13}}},\\[2mm]
Z =\frac{|x_{23}||x_{13}|}{|x_{12}|} \left(\frac{x_{13}^\mu}{x_{13}^2}-\frac{x_{23}^\mu}{x_{23}^2}\right) \zeta_\mu\ .
\end{gather}
In more general cases, the number $N_3$ of tensor structures can be computed in terms of the
representation theory of the rotation group \cite{Mack:1976pa,Osborn:1993cr}
\begin{equation} \label{N3}
N_3(\mu_1,\mu_2,\mu_3) = \sum_\mu N_{\mu_1,\mu'_2}^\mu n_\mu(\mu_3) \ ,
\end{equation}
where the sum runs over irreducible representations $\mu$ of the rotation group and
$N_{\mu_1,\mu'_2}^\mu$ denotes the Clebsch-Gordon multiplicities for the decomposition
of the tensor product of $\mu_1$ and $\mu'_2$. The number $n_\mu$ denotes the number of
\(\SO(d-1)\)$\subset $ \(\SO(d)\) invariant linear maps from \(V_\mu\) to
\(V_{\bar{3}}\) i.e.
\begin{gather}\label{nmu}
 n_\mu(\mu_3) = \textit{dim}\left( \textit{Hom}_{\SO(d-1)}(V_\mu , V_{\bar{3}} )\right)\ .
\end{gather}
Here $V_\mu$ and $V_{\bar{3}}$ are the carrier spaces of the representations $\mu$
and $\bar{\mu}_3$, respectively. The subscript indicates that we consider only \(\SO(d-1)\)
invariant maps. Let us note that the number $N_3$ of 3-point tensor structures $t_{123}$
also counts the number of different tensor structures appearing in the operator product
expansion of the first two fields \(\mathcal{O}_1\) and \(\mathcal{O}_2\) into the third \(\mathcal{O}^\dag_3\).
From our description it is clear that we can construct the tensor structures in operator products as
$t_{123} = \sum_\mu I_{\mu\bar{\mu}_3}C_{12'\mu}$. Here, $C_{12'\mu}$ is a $\SO(d)$
Clebsch-Gordon map from the tensor product $\mu_1 \otimes \mu'_2$ into the $\SO(d)$
representations $\mu$. The maps $I_{\mu\bar{\mu}_3}$, on the other hand, are $\SO(d-1)$
intertwiners between the representations $\mu$ and $\bar{\mu}_3$ where both are
restricted to representations of the subgroup $\SO(d-1) \subset \SO(d)$.

Even though formula \eqref{N3} seems to break the symmetry between $1,2,3$, the number
it computes is actually completely symmetric. In fact, inserting eq.\ \eqref{nmu} into eq.
\eqref{N3} we obtain
\begin{equation}\label{N3b}
N_3(\mu_1,\mu_2,\mu_3) = \textit{dim} \left( \textit{Hom}_{\SO(d-1)}(V_1\otimes V_{2'},
V_{\bar{3}})\right) =  \textit{dim} \left(V_{1}\otimes V_{2'} \otimes V_3\right)^{\SO(d-1)}\ .
\end{equation}
The relevance of the subgroup $\SO(d-1) \subset \SO(d)$ is not too difficult to understand.
Recall that we can use conformal transformations to move three points in $\mathbb{R}^d$ to
the origin, the point $e_1 = (1,0,\dots,0)$ and the point at infinity. Since all these
points lie on a single line $\mathbb{R} \subset \mathbb{R}^{d}$, the configuration is
left invariant by rotations of the transverse space $\mathbb{R}^{d-1}$.
\medskip

After this preparation let us turn to the main object of our interest, namely the 4-point correlation function. Similarly to the case of 2 and 3-point correlation functions it can
be decomposed into the sum over different tensor structures \(t^I=t^I_{1234}\)
\begin{gather}\label{4pTS}
\langle\mathcal{O}_1(x_1)\mathcal{O}_2(x_2)\mathcal{O}_3(x_3)\mathcal{O}_4(x_4)\rangle=
\Omega_{(12)(34)}(x_i)\sum\limits_{I=1}^{N_4}g^I(u,v)t^I_{1234},\\ \notag
\Omega_{(12)(34)}(x_i) = \frac{1}{x_{12}^{\D_1+\D_2}x_{34}^{\D_3+\D_4}}
\left(\frac{x_{14}}{x_{24}}\right)^{\D_2-\D_1}
\left(\frac{x_{14}}{x_{13}}\right)^{\D_3-\D_4}\ .
\end{gather}
The coefficients \(g^{I}(u,v)\) depend on two anharmonic ratios \(u= x^2_{12}x^2_{34}/ x^2_{13}x^2_{24}\) and \( v= x^2_{14}x^2_{23}/x^2_{13}x^2_{24}\) and \(N_4\) is the
number of different 4-point tensor structures,
\begin{equation}
N_4 \equiv N_4(\mu_1,\mu_2,\mu_3,\mu_4) \leq
\textit{dim} \left(V_{1}\otimes V_{2'} \otimes V_3\otimes V_{4'}\right)^{\SO(d-2)}\ .
\end{equation}
This formula is a direct extension of formula \eqref{N3b} for the number of 3-point
structures. The main difference is that now we need to look for invariants with respect
to the action of $\SO(d-2) \subset \SO(d)$ rather than $\SO(d-1)$. Once again, we can
understand the relevance of this subgroup from the geometry of insertion points in
$\mathbb{R}^d$. It is well known that conformal transformations allow to bring four
such points into a 2-dimensional plane $\mathbb{R}^2\subset \mathbb{R}^d$. The
subgroup $\SO(d-2)$ is the symmetry group of the associated transverse space.

As in our analysis of 3-point structures, we obtain an alternative view on the
tensor structures if we evaluate 4-point correlation functions by performing
operator product expansion of two fields $\mathcal{O}_1$ and $\mathcal{O}_2$
into conformal primary fields $\mathcal{O} = \mathcal{O}_\pi$ and its
descendants. The result reads as
\begin{gather}
\langle\mathcal{O}_1(x_1)\mathcal{O}_2(x_2)\mathcal{O}_3(x_3)\mathcal{O}_4(x_4)\rangle
=\sum \limits_{\mathcal{O}_\pi}\sum\limits_{\alpha,\beta}\lambda^\alpha_{12\pi}\lambda^\beta_{\bar{\pi} 34}
W^{\alpha\beta}_{1234,\pi}(x_1,x_2,x_3,x_4) \ . \label{4pOverCPW}
\end{gather}
The set of 3-point tensor structures \(\alpha,\beta\) that appear in the
two operator products depends on the intermediate operator \(\mathcal{O}_\pi\)
with $\pi = (\Delta,\mu)$. The individual block $W$ may now be decomposed as
\begin{gather}\label{CPWoverCB}
W^{\alpha\beta}_{1234,\pi}(x_1,x_2,x_3,x_4) =
\Omega_{(12)(34)}(x_i) \, \sum\limits_I g^{I,\alpha\beta}_{\Delta,\mu}(u,v)
\, t^{I}_{1234}\ ,
\end{gather}
It is clear that not all 4-pt tensor structures appear in the decomposition (\ref{CPWoverCB}) :
 \begin{equation}\label{N4}
N_4(\mu,\alpha,\beta) =  N_4(\mu_1,\mu_2,\mu_3,\mu_4;\mu,\alpha,\beta) \leq N_4.
 \end{equation}
We can now perform the decompositions \eqref{4pOverCPW} and \eqref{CPWoverCB}
on the coefficients $g^I(u,v)$ defined in eq.\ \eqref{4pTS} to obtain the
following expansion in terms of spinning conformal blocks
$g^{I,\alpha\beta}_\pi (u,v)$,
\begin{gather}\label{OverCB}
g^I(u,v)=\sum \limits_{\mathcal{O}_\pi}\sum\limits_{\alpha,\beta}\lambda^\alpha_{12\pi}\lambda^\beta_{\bar{\pi} 34}g^{I,\alpha\beta}_{\Delta,\mu}(u,v)\ .
\end{gather}
The spinning conformal blocks \([g^{I,\alpha\beta}_{\Delta,\mu}(u,v)]\) with given
\(\pi = (\Delta, \mu)\) satisfy a set of second order differential equations of the
form
\begin{gather}
\mathcal{C}^{(2)} [g^{I',\alpha'\beta'}_{\Delta,\mu}(u,v)] = C_{\Delta,\mu}\ [g^{I,\alpha\beta}_{\Delta,\mu}(u,v)]\ ,
\end{gather}
where $\mathcal{C}^{(2)}$ denotes the second order Casimir differential operator and
$C_{\Delta,\mu}$ is the eigenvalue of the quadratic Casimir element of the conformal
group in the representation $\chi_\pi$ that is induced from $(\Delta,\mu)$. Such
Casimir equations are well known for scalar blocks, see \cite{Dolan:2003hv}, and they
were constructed for several examples involving fields with spin, see
\cite{Iliesiu:2015akf,Echeverri:2016dun}. Our main goal in this work is to develop
a systematic approach to Casimir equations for spinning blocks.

\section{Harmonic Analysis Approach to Conformal Blocks }

In the previous section we described spinning conformal blocks as a set
of functions $g^{\alpha \beta}_{\Delta,\mu}(u,v)$ of the two anharmonic
rations one can build out of four points in $\mathbb{R}^d$. The main goal
of the current section is to show that the same objects can also be realized
as sections of a certain vector bundle over a 2-dimensional quotient
of the conformal group $G$ itself. While our discussion will remain a bit
abstract, it mirrors the line of arguments we went through in the previous
section. Many of the key elements will be illustrated in the next section
when we discuss concrete examples.

In mathematical terms, 4-point conformal blocks are invariants in the tensor
product of four continuous series representations $\chi_i, i=1, \dots,4$ of
the conformal group $G$. In the principal continuous series, the conformal
weights are of the form $\Delta = d/2+ic$ with real parameter $c$.
We shall adopt these values for now and only continue to real conformal
weights at the very end once we derived the equations. In order to construct
this space, we will first realize the tensor products \(\chi_1\otimes \chi_2\)
and \(\chi_3\otimes \chi_4\) in a space of functions on $G$ with certain
equivariance properties under the left/right regular action of the subgroup
$K \subset G$. According to theorem 9.2 of \cite{Dobrev:1977qv} the tensor
product $\chi_{\pi_1} \otimes \chi_{\pi_2}$ can be realized as
\begin{eqnarray}
  \label{TPLC}
  \chi_{\pi_1} \otimes \chi_{\pi_2} & \cong &
  \Gamma^{(\pi_1,\pi_2)}_{K\setminus G} \quad
  \textrm{ with } \\[2mm]  \nonumber
  \Gamma^{(\pi_1,\pi_2)}_{K\setminus G}
 &=&  \left\{
f:g \rightarrow V_{\mu_1} \otimes V_{\mu'_2}\,
\Biggr\rvert  \,
\begin{array}{ll}
 f(d(\lambda)g) = e^{\lambda(\Delta_2-\Delta_1)} f(g) \quad &
\mbox{ for } \ d(\lambda) \in D \subset G \\[2mm]
 f(rg) = \mu_1(r) \otimes \mu'_2(r) f(g) & \mbox{ for }
r \in R \subset G \end{array} \right\} .
\end{eqnarray}
Here, $V_{\mu_1}$ and $V_{\mu'_2}$ denote the finite dimensional carrier
spaces of our representations $\mu_1$ and $\mu'_2$ of the rotation group
and we wrote elements $d \in D$ as
\begin{equation}
d(\lambda) = \left( \begin{array}{cc} \cosh\lambda & \sinh\lambda \\
\sinh\lambda & \cosh\lambda \end{array} \right)\ .
\end{equation}
For a proof of this theorem see \cite{Dobrev:1977qv}. Elements of the space \eqref{TPLC}
are vector valued functions on the group that are covariantly constant along
the orbits of the left $K$-action on $G$. Such functions are uniquely characterized
by the values they assume on the space $K\setminus G$ of such orbits. This is why we
shall often refer to $\Gamma$ as a space of sections in a vector bundle over
the quotient space $K\setminus G$. Similarly one can realise tensor product
\(\chi_{\pi_3} \otimes \chi_{\pi_4}\) on the right cosets \(G/K\),
\begin{eqnarray}
  \label{TPRC}
  \chi_{\pi_3} \otimes \chi_{\pi_4} & = &  \Gamma^{(\pi_3,\pi_4)}_{G/K} \quad
  \textrm{ with } \\[2mm]  \nonumber
  \Gamma^{(\pi_3,\pi_4)}_{G/K}
 &=&  \left\{
f:g \rightarrow V_{\mu_3} \otimes V_{\mu'_4}\,
\Biggr\rvert  \,
\begin{array}{ll}
 f(gd(\lambda)^{-1}) = e^{\lambda(\Delta_4-\Delta_3)} f(g) \quad &
\mbox{ for } \ d(\lambda) \in D \subset G \\[2mm]
 f(gr^{-1}) = \mu_3(r) \otimes \mu'_4(r) f(g) & \mbox{ for }
r \in R \subset G \end{array} \right\} .
\end{eqnarray}
Let us note in passing that the spaces $\Gamma$ we defined in eqs. \eqref{TPLC}
and \eqref{TPRC} decompose into an infinite set of irreducible representations of
the conformal group. The number of times a given representation $\chi_\pi = \chi_3$
appears in this decomposition is given by the formula \eqref{N3b} for the number of
3-point tensor structures.
\medskip

Equipped with a good model for the tensor products of field multiplets we now want
to realize $G$-invariants in the four-fold tensor product of representations. In
order to keep the discussion as transparent as possible we shall first restrict to
the case of four external scalars, i.e.\ we shall assume that $\pi_i = (\Delta_i,
\mu_i)$ with $\mu_i = 0$. As before, we group these four fields into two pairs and
apply the previous theorem to realize the products of representations $\chi_1
\otimes \chi_2$ and  $\chi_3 \otimes \chi_4$ on the vector bundles
\eqref{TPLC} and \eqref{TPRC},
respectively. Since these bundles are defined over the left and right cosets $K
\setminus G$ and \(G/K\), respectively, they both carry an action of the conformal
group $G$ by right resp. left translations. More precisely, an element \(g\in G\)
acts on \(K\setminus G\times G/H\) as \((g_1,g_2)\rightarrow (g_1 g^{-1},g g_2)\).
We can use this action to pass to the space of invariants,
\begin{eqnarray} \label{s4ptHA}
  \left( \bigotimes_{i=1}^4 \chi_{\pi_i} \right)^G & \cong &
  \left(\Gamma^{(\Delta_1,\Delta_2)}_{K\setminus G} \otimes
\Gamma^{(\Delta_3,\Delta_4)}_{G/K}\right)^G \ \cong \
\Gamma^{(a,b)}_{G\dcoset K}  \quad \textrm{ with } \\[2mm]
\Gamma^{(a,b)}_{G\dcoset K}& = &\left\{ f:G\rightarrow \mathbb{C} \bigr\rvert
f(d(\lambda)g) = e^{2a\lambda} f(g)\ , \ f(gd(\lambda)^{-1})= e^{2b\lambda} f(g)\right\}\
\end{eqnarray}
where $2a =\Delta_2-\Delta_1$ and $2b = \Delta_3-\Delta_4$. Since we have assumed
that $\Delta_i = d/2 + ic_i$, the parameters $a,b$ are purely imaginary before we
continue to real $\Delta_i$. We have now obtained
a new model for the space of conformal blocks \(g(u,v)\). Since we restricted to
four external scalars, there is a single tensor structure only so that no indices
$I,\alpha\beta$ appear. In our notations we indicate that we want to think of the
space \eqref{s4ptHA}, as a space of sections in a line bundle over the double coset
$G\dcoset K$. The latter appears since $(K\setminus G \times G/K)/G = K\setminus G/K
\equiv G\dcoset K$ which follows from the obvious relation $(G\times G)/G = G$. As
we will see in the next section, the double coset $G\dcoset K$ is two-dimensional
and the two coordinates are related with the two anharmonic ratios \(u, v\)
we used  in the previous section. In complete analogy with the decomposition (\ref{OverCB})
we can decompose the space \(\Gamma^{(a,b)}_{G\dcoset K}\) of sections into a sum
over intermediate channels,
\begin{gather}\label{FiberOverScBks}
\Gamma^{(a,b)}_{G\dcoset K}=\bigoplus\limits_{\Delta,\mu}\
\Gamma^{(a,b),(\Delta,\mu)}_{G\dcoset K}\ .
\end{gather}
Since we constructed $\Gamma$ as a space of functions on $G$ with certain
equivariance properties, the Laplacian on the conformal group $G$ descends to $\Gamma$
and the decomposition \eqref{FiberOverScBks} is the corresponding spectral decomposition.
More precisely, the summands in the decomposition are eigenspaces of the Laplacian with
eigenvalue $C_{\Delta,\mu}$ and certain boundary conditions.
\smallskip

It remains to extend the previous discussion to the case of spinning blocks, i.e. we
need to drop the condition $\mu_i = 0$. Formula \eqref{s4ptHA} possesses the following
extension to cases with $\mu_i \neq 0$,
\begin{equation} \label{4ptHA}
\left(\Gamma^{(\Delta_1,\mu_1;\Delta_2,\mu_2)}_{K\setminus G} \otimes
\Gamma^{(\Delta_3,\mu_3;\Delta_4,\mu_4)}_{G/K}\right)^G \cong
\Gamma^{(a,\mu_1\otimes\mu'_2;-b,\mu_3\otimes \mu'_4)}_{G\dcoset K} \ .
\end{equation}
The labels $a,b$ are determined by the conformal weights of the external fields as before.
Extending our prescription \eqref{s4ptHA}, we specify vector bundle over $G\dcoset K$ that
appear on the right hand side in the following way
\begin{equation} \label{sbundles}
  \Gamma^{(\L\R)}_{G\dcoset K} = \{\, f:G \rightarrow V_{\L} \otimes
  V_{\R} \, | \, f(kg)= \L(k) f(g) \quad ,\quad f(gk^{-1}) = \R(k)
  f(g) \ \}\ ,
\end{equation}
where the two representations \(\L=(a,\mu_1\otimes\mu'_2)\), \(\R=(-b,\mu_3\otimes\mu'_4)\)
act on \(V_{\L}= V_1 \otimes V_{2'} \) and \(V_{\R}= V_3 \otimes V_{4'}\), respectively,
according to
\begin{gather}
\L(d(\lambda)r)=e^{2a\lambda}\mu_1(r)\otimes \mu'_2(r)\ \ ,\ \ \
\R(d(\lambda)r)=e^{-2b\lambda}\mu_3\otimes \mu'_4(r)\ .
\end{gather}
Our definition \eqref{sbundles} selects a subspace among functions on the group that
take values in the 4-fold tensor product $V_{\mu_1} \otimes V_{\mu'_2} \otimes
       V_{\mu_3} \otimes V_{\mu'_4}$ of the group
$K$. The identification of this space as sections of a vector bundle over the coset
space is a bit more tricky in $d>3$ since the action of $K \times K$ on the conformal
group $G$ is not free beyond $d = 3$ dimensions. As we shall see
explicitly in the next section, the stabilizer for the action of $K \times K$ on $G$
is given by a subgroup $\SO(d-2)\subset \SO(d) \times \SO(d)$. If we now want to construct a
function $f$ in the space \eqref{sbundles} by prescribing the values it takes on the
double coset, we have to make sure that the covariance conditions with respect to the
left and right action of $K$ are compatible. This compatibility condition forces $f$
to take values in the
subspace
\begin{equation} \label{fiber}
 T = \left( V_{\mu_1} \otimes V_{\mu'_2} \otimes
       V_{\mu_3} \otimes V_{\mu'_4}\right)^{\SO(d-2)}\ .
\end{equation}
In conclusion, we can indeed think of the space \eqref{sbundles} as a space of
sections in a vector bundle over the double coset, as long as we remember to restrict
the fibers to the space of $\SO(d-2)$ invariants in the tensor product of the spin
representations. Note that the space $T$ contains the space of 4-point tensor
structures we introduced in the previous section.

As in eq.\ (\ref{FiberOverScBks}) we can decompose the space (\ref{4ptHA}) into a
sum of eigenspaces of the Laplacian of the conformal group,
\begin{gather}\label{blockdec}
\Gamma^{(\L\R)}_{G\dcoset K}=\sum\limits_{\Delta,\mu}\Gamma^{(\L\R),(\Delta,\mu)}_{G\dcoset K}\ .
\end{gather}
We have now succeeded to model spinning conformal blocks through vector
valued $K \times K$-covariant functions on $G$. The latter can also be thought
of as sections in vector bundles over the double coset $G\dcoset K$. Our next
task is to analyse the action of the Laplacian on the spaces \eqref{blockdec}
and finally to compare the associated eigenvalue problem with the Casimir
equations for conformal blocks.

\section{Harmonic Analysis and Calogero-Sutherland Models}

Our goal in this section is to describe an algorithm that allows us to write
the action of the conformal Laplacian on the spaces \eqref{sbundles} as a
Hamiltonian for two interacting particles with spin that move on a
1-dimensional space. The latter will turn out to be of Calogero-Sutherland
type. This extends a classical observation of Olshanetsky and Perelomov
about a relation between certain harmonic analysis problems on groups and
Calogero-Sutherland Hamiltonians \cite{Olshanetsky:1976,Olshanetsky:1981,
Olshanetsky:1983} to the cases with spin, see also \cite{Feher:2007ooa,
Feher:2009wp}. In the context of conformal field theory, our findings
generalize \cite{Isachenkov:2016gim} to spinning blocks.

In order to achieve our goal we shall introduce a special set of coordinates on
the conformal group that are based on a variant of the Cartan decomposition and suited for identification of double quotient $G//K$,
see first subsection. We will then construct the Laplace-Beltrami operator
on the conformal group in these coordinates. In a final step, we integrate
over the $K\times K$ orbits to obtain second order differential operators
on the 2-dimensional quotient space $G\dcoset K$. The latter can be
transformed into a Calogero-Sutherland type Hamiltonian.

\subsection{Cartan decomposition of the conformal group}

Let us begin by introducing a coordinate system on the conformal group $G=
\SO(1,d+1)$ that is well adapted to the action of the $K \times K \subset
G \times G$ on $G$. The action of $G \times G$ that we restrict to the subgroup
$K\times K$ is the action on $G$ by left and right regular translations. Our
choice of $K = \SO(1,1) \times \SO(d)$ determines a so-called Cartan or KAK
decomposition of $G$. In order to describe the details we note that Lie algebra
$\mathfrak{k}$ of \(K\) contains all the elements of Lie algebra $\mathfrak{g}$ of conformal group \(G\) that are eigenvectors with eigenvalue \(+1\) of automorphism \(\Theta\) acting on \(\xi\in\mathfrak{g}\) as \(\Theta(\xi)=\theta \xi \theta,\ \theta =\, $\textit{diag\/} $(-1,-1,1,\dots,1)\). The automorphism $\Theta$
determines a decomposition of the Lie algebra $\mathfrak{g}$ of the conformal
group $G$ as $\mathfrak{g} = \mathfrak{k} \oplus \mathfrak{p}$ where $\mathfrak{k}$
is the Lie algebra of the subgroup $K$ and $\mathfrak{p}$ its orthogonal complement.
The latter is the subspace on which $\Theta$ acts by multiplication with $-1$.
What leads to \(\mathbb{Z}_2\) grading on $\mathfrak{g}$
$$ [\mathfrak{k}, \mathfrak{k}] \subset \mathfrak{k} \quad , \quad
  [\mathfrak{k}, \mathfrak{p}] \subset \mathfrak{p} \quad , \quad
[\mathfrak{p}, \mathfrak{p}] \subset \mathfrak{k}\ . $$
Note that any \textit{subalgebra} $\mathfrak{a}\subset \mathfrak{p}$ of $\mathfrak{g}$ that is
contained in $\mathfrak{p}$ must be abelian. Choosing a certain 2-dimensional subalgebra
\(\mathfrak{a}\) and then exponentiating it we get the abelian subgroup \(A\) and the Cartan decomposition
reads as \(G=KAK\).\footnote{Let us stress that the decomposition $g = k_1 a k_2$
of an element $g \in G$, if it exists, is not unique for $d > 3$, see below.}

Now let us describe the Cartan decomposition explicitly. To this end, we shall work
with the usual set of generators $M_{ij}= - M_{ji}$ of the conformal group $G=$
$\SO(1,d+1)$ where $i,j$ run through $i,j= 0,1,2, \dots d+1$. Here
$i,j=0$ correspond to the time-like direction while all other
directions are space-like. Obviously, the Lie algebra $\mathfrak{k}$
of $K$ is spanned by the generator $M_{0,1}$ of dilations along with
the elements $M_{\mu\nu}$ for $\mu = 2, \dots,d+1$ that generate
rotations. Our subspace $\mathfrak{p}$ in turn is spanned by
$M_{0,\mu}$ and $M_{1,\mu}$. The choice of $\mathfrak{a}$ that
we shall adopt is the one for which $\mathfrak{a}$ is spanned by
$a_+ = M_{0,2}$ and $a_- = M_{1,3}$. These two generators commute
with each other since they have no index in common. Clearly, the
Cartan algebra cannot be extended beyond $a_+,a_-$ since any other
generator $\mathfrak{p}$ will necessarily have one index in common
with the ones we have singled out as $a_+$ and $a_-$.

Through the Cartan decomposition we may write any element $g \in G$
of the conformal group as a product of the form
\begin{equation} \label{KAK}
  g = d(\lambda_1)\, r_1 \, a(\tau_1,\tau_2) \, d(\lambda_2) \, r_2  .
\end{equation}
Here $d(\lambda_i) \in D=$ \(\SO(1,1)\) are considered as elements of
the subgroup $D \subset G$. The group element $a(\tau_1,\tau_2)$ in
turn is given by
\begin{equation}
\label{adef}
a(\tau_1,\tau_2) =  \left( \begin{array}{cccc}
\cosh\frac{\tau_1}{2} & 0 & \sinh\frac{\tau_1}{2} & 0 \\
0 & \cos\frac{\tau_2}{2} & 0 & -\sin\frac{\tau_2}{2} \\
\sinh\frac{\tau_1}{2} & 0 & \cosh\frac{\tau_1}{2} & 0 \\
0 & \sin\frac{\tau_2}{2} & 0 & \cos\frac{\tau_2}{2}
\end{array} \right)   .
\end{equation}
There is one small subtlety that is associated with elements $r_1$
and $r_2$ of the rotation group. Let us note that the two generators
$a_+$ and $a_-$ of our subgroup $A \subset G$ are left invariant by
all generators of the form $M_{\mu\nu}\in \mathfrak{k}$ with $\mu,\nu
=4,\dots,d+1$. These generate a subgroup $B = \SO(d-2) \subset$
\(\SO(d)\subset K\). Consequently, the decomposition \eqref{KAK} is not
unique as we can move factors $b \in B$ between $r_1$ and $r_2$. We
can fix this freedom by choosing $r_2$ to be a representative of a
point on the coset space $K/B=$ $\SO(d)/$ $\SO(d-2)$. Once this choice
is adopted, the KAK decomposition becomes unique up to discrete
identifications. One may verify that the dimensions indeed match
$$\mbox{\it dim\/} G = \mbox{\it dim\/}K + \mbox{\it dim\/}A +
\mbox{\it dim\/}K - \mbox{\it dim\/}B\ .$$
To complete our description of coordinates on the conformal group it
remains to parametrize the elements $r_i$ of the rotation group. The detailed
choice does not matter since these coordinates will be integrated
over later.

In the remainder of this work we shall assume that $d\leq 3$ so that the group
\(B\) is trivial. Extending our calculations beyond $d=3$ is the subject of a
future paper \cite{SS}.
\bigskip

\noindent
\textbf{Example:} Throughout this section we shall illustrate all
our statements and constructions at the example of the 2-dimensional
conformal group $\SO(1,3)$. In this case we shall parametrize the
elements $r_1=r_1(\phi_1)$ and  $r_2=r_2(\phi_2)$ such that
$$
k_i(\lambda_i,\phi_i) = d(\lambda_i) r(\phi_i) = \left(
\begin{array}{cccc}
 \cosh \lambda_i & \sinh \lambda_i & 0 & 0 \\
 \sinh \lambda_i & \cosh \lambda_i & 0 & 0 \\
 0 & 0 & \cos \phi_i & -\sin \phi_i \\
 0 & 0 & \sin \phi_i & \cos \phi_i \\
\end{array}
\right)\ .
$$
Thereby we have now parametrized an arbitrary element of the conformal
group $\SO(1,3)$ with the help of the product formula \eqref{KAK} through
the six coordinates $\lambda_i,\phi_i,\tau_i$ for $i=1,2$. These coordinates
possess the following ranges: $\tau_1\in(-\infty,\infty),\ \tau_2\in[0,4\pi)$,
$\lambda_i\in(-\infty,\infty)$, $\phi_i\in[0,2\pi)$.

\subsection{The Laplacian on the Cartan subgroup}

Our next task is to construct the Laplacian on the conformal group in
the coordinate system we have introduced in the previous subsection.
This is straightforward. The Laplace-Beltrami operator on any
Riemannian manifold may be computed from the metric $g$ through
\begin{gather}
  \Delta_{\textrm{LB}} = \sum\limits_{\alpha,\beta} |\det (g_{\alpha\beta})|^{-\frac{1}{2}}
  \partial_{\alpha} g^{\alpha \beta}|\det (g_{\alpha\beta})|^{\frac{1}{2}}
  \partial_\beta \ . \label{LBdef}
\end{gather}
On a group manifold the metric \(g_{\alpha\beta}\) is obtained with the
help of the Killing form as
\begin{gather}
  g_{\alpha\beta}(x)=-2\ \text{tr}\ h^{-1}\partial_\alpha h \
  h^{-1}\partial_\beta h, \ \ \ h\in G \ .
\end{gather}
By construction, the Laplace-Beltrami operator $\Delta_{\textrm{LB}}$ commutes with
the $G \times G$ action on the group $G$ by left and right regular transformations.
Since it is a second order differential operator, it can be written as a quadratic
expression in the left or right invariant vector fields on $G$ in which the vector
fields are contracted with the Killing form, i.e.\ the Laplace-Beltrami operator
coincides with the action of the quadratic Casimir element on functions.

In the setup we described in the previous section, the Laplace-Beltrami operator
acts on functions $f$ on the conformal group that take values in the vector spaces
$V_\L \otimes V_\R$. Since the bundle over the group $G$ is trivial, the Laplace
operator acts simply component-wise. We will not distinguish in notation between
the Laplacian on the group itself and on trivial vector bundles.

Using the metric on $G$ we can also construct the invariant Haar measure $d\mu_G$
on $G$. Its density is given by $\sqrt{\det g_{\alpha,\beta}}$. The Haar measure can
then be used to introduce a scalar product for (vector-valued) functions on $G$.
The associated space of square integrable functions will be denoted as usual by
$L_G^2=L^2(G,V_\L \otimes V_\R;d\mu_G)$. The Laplace-Beltrami operator is a densely
defined on this space and it is Hermitian with respect to the scalar product.
\bigskip

\noindent
\textbf{Example:} Using the coordinates on $\SO(1,3)$ that we introduced at the
end of the previous subsection, the metric takes the form
\begin{eqnarray}
  g_{\alpha\beta}dx^\alpha dx^\beta
  &=& 4(d^2\phi_l+d^2\phi_r-d^2\lambda_l-d^2\lambda_r)-d^2\tau_1+d^2\tau_2\notag\\[2mm]
  & & -\hspace*{-2cm}
  -8\sinh\frac{\tau_1}{2}\sin\frac{\tau_2}{2}(d\lambda_ld\phi_r+d\lambda_rd\phi_l)+
8\cosh\frac{\tau_1}{2}\cos\frac{\tau_2}{2}(d\phi_l d\phi_r-d\lambda_l d\lambda_r)\ .
\end{eqnarray}
It is easy to work out the Haar measure on the conformal group from the determinant
of the metric,
$$ d\mu_G=8(\cosh\tau_1-\cos\tau_2)d\lambda_l d\phi_l d\tau_1d\tau_2 d\lambda_r d\phi_r\ .
$$
We leave it as an exercise to construct the associated Laplace-Beltrami operator.
\medskip

In the context of d-dimensional conformal blocks we are now instructed to
restrict the Laplace-Beltrami operator to the space \eqref{sbundles} and to
study the spectrum and eigenfunctions of this restriction. The elements of
the space \eqref{sbundles} are $K\times K$ covariant functions on the group
$G$ and hence they are uniquely characterized by their dependence on the two
coordinates $\tau_1$ and $\tau_2$. We can equip functions in the Cartan
subgroup $A$ with a measure\footnote{Recall that we assumed $d\leq 3$. For
$d\geq 4$, the integration is over fibers of the $K\times K$ action on $G$.}
$$ m(\tau_1,\tau_2) d\tau_1d\tau_2 := d\mu_A(\tau_1,\tau_2) \:= \frac{1}{Z}
\int^\textrm{reg}_{K\times K} d\mu_G $$
with $Z = \textit{Vol}(\SO(d))^2$. Note that $K= D \times R$ contains the
non-compact factor $D$ that makes the integration over $K$ divergent. We
can regularize this divergence e.g. through the prescription
$$ \int^\textrm{reg}_\mathbb{R} d\lambda = \lim_{L\rightarrow \infty}
\frac{1}{2L} \int_{-L}^L d\lambda \ . $$
Having fixed a measure on $A$ we can now take a function $f_A\in L^2_A=
L^2(A,V_\L \otimes V_\R;d\mu_A)$ on the Cartan subgroup $A$ with values
in the linear space $V_\L \otimes V_\R$. Such a function may be extended
uniquely to a $V_\L \otimes V_\R$-valued covariantly constant function
on $G$. The latter is square integrable provided we agree to regularize
the integration over $\lambda_l$ and $\lambda_r$ as outlined above. In
other words, there is an isomorphism of Hilbert spaces
\begin{gather}
 L^2(\Gamma^{(\L\R)};d\mu_G) \cong
L^2_A = L^2(A,V_\L \otimes V_\R;d\mu_A)\ .
\end{gather}
This isomorphism induces a correspondence between $K\times K$
invariant Hermitian differential operators $\mathcal{D}$ acting on $L^2_G$ and
Hermitian differential operators $\mathcal{D}^A$ on the Cartan subgroup $A$
such that
\begin{eqnarray}
\int_A d\mu_A\langle f_A,\mathcal{D}^A g_A\rangle & = &
 \frac{1}{Z}\int^\textrm{reg}_G d\mu_G \langle f(k_lak_r),\mathcal{D}\,  g(k_lak_r)\rangle,\label{Reduction}\\[4mm]
\textrm{where} \quad  f(k_lak_r) & = & [\mathcal{L}(k_l)\otimes\mathcal{R}(k_r^{-1})]
f_A(\tau_1,\tau_2)=\notag\\[2mm] & = & e^{2a\lambda_l+2b\lambda_r}[(\mu_1\otimes\mu'_2)(r_l)
\otimes(\mu_3\otimes\mu'_4)(r_r^{-1})]
f_A(\tau_1,\tau_2)\ . \notag
\end{eqnarray}
Here, $f$ and $g$ are two covariantly constant functions on $G$, i.e. two
elements of the space \eqref{sbundles}. The symbols $f_A$ and $g_A$ denote
their restriction to the Cartan subgroup $A \subset G$. Elements \(k_l,\ k_r\)
are parametrized as \(k_l=d(\lambda_l)r_l,\ k_r=d(\lambda_r)r_r\). In addition
we used $\langle \cdot,\cdot \rangle$ for the scalar product on the finite
dimensional linear space $V_\L \otimes V_\R.$%
\medskip

We can now apply the prescription \eqref{Reduction} to the Laplacian $\mathcal{D}
= \Delta_\textrm{LB}$. In order to bring the reduced Laplacian $\Delta_\textrm{LB}^A$
into the form of a Calogero-Sutherland Hamiltonian on a space with measure $d\tau_1
 d\tau_2$, it remains to remove the non-trivial factor $m(\tau_1,\tau_2)$ in the
measure on the Cartan subgroup by an appropriate gauge transformation. This is achieved
by rescaling the functions \(f_A\in L^2_A\) such that
$$\psi_A(\tau_1,\tau_2) = \sqrt{m(\tau_1,\tau_2)}\,  f_A(\tau_1,\tau_2) \ . $$
On the 2-particle wave functions $\psi_A(\tau_1,\tau_2)$ the reduced Laplacian indeed
takes the form of a Calogero-Sutherland type Hamiltonian,
\begin{gather}\label{LBAtoHam}
  H_{(\L,\R)} \ = \ \sqrt{m(\tau_1,\tau_2)}\, \Delta^{A}\, \frac{1}{\sqrt{m(\tau_1,\tau_2)}}
  =: -\frac{d^2}{d\tau^2_1} + \frac{d^2}{d\tau^2_2} + V_{(\L,\R)}(\tau_1,\tau_2) \ .
\end{gather}
After performing the gauge transformation that trivialized the measure, we can read off
the matrix valued potential $V_{(\L,\R)}$. It depends on the choice of the representations
$\L,\R$ and acts on the space $V_\L \otimes V_\R$. Our construction guarantees that the
Hamiltonian $H_{(\L,\R)}$ is Hermitian with respect to the measure $d\tau_1d\tau_2$ as it
descends from the Hermitian Laplace-Beltrami operator on the conformal group $G$.
In conclusion, we have now described an algorithm that associates a family of
matrix valued potentials $V_{(\L,\R)} = V_{(a,\mu_1\otimes\mu'_2;-b,\mu_3\otimes\mu'_4)}$ to
any spinning conformal block. In order to make the kinetic term of the model look more
standard, we will often use the coordinates $\tau_1=x+y$ and $\tau_2 =i(x-y)$.
\medskip

\noindent \textbf{Example:} Returning to our example of $G=\SO(1,3)$ we want to determine
the action of the Laplace-Beltrami operator on scalar blocks. In the case of scalars with
parameters $a,b$, the covariantly constant functions on $G$ read
$$ f(x) = e^{2a\lambda_l+2b\lambda_r} f_A(\tau_1,\tau_2)\ . $$
Our reduction formula (\ref{Reduction}) for the Laplacian becomes
\begin{eqnarray}
\int d\mu_A \bar f_A(\tau_1,\tau_2) \Delta^A g_A(\tau_1,\tau_2) & = &\notag\\[2mm]
& & \hspace*{-5cm} = \int d\tau_1 d\tau_2 \,  (\cosh\tau_1-\cos\tau_2) e^{-2a\lambda_l-2b\lambda_r}
\bar f_A(\tau_1,\tau_2) \Delta_{\textrm{LB}}\left(e^{2a\lambda_l+2b\lambda_r}
g_A(\tau_1,\tau_2)\right) \ .
\end{eqnarray}
Here, $\bar f_A$ is the complex conjugate and we have used that $a$ and $b$ are purely
imaginary. The measure $d\mu_A$ on $A$ is given by $d\mu_A = m d\tau_1d\tau_2$ with a non-trivial
density function $m(\tau_1,\tau_2)=\cosh\tau_1-\cos\tau_2$. If we perform the transformation
(\ref{LBAtoHam}) with the square root $m=(\cosh\tau_1-\cos\tau_2)^{\frac{1}{2}}$ of the
measure factor we obtain the famous Calogero-Sutherland Hamiltonian of \BC\ type
\begin{gather}
 H =  \frac{1}{2}H^{(a,b,0)}_{C.S}+\frac{1}{4}
\end{gather}
where
\begin{gather}\label{sCSHam}
H^{(a,b,\epsilon)}_{CS}=-\partial^2_x-\partial^2_y+V^{(a,b,\epsilon)}_{C.S.}, \ \ \ \ \epsilon=d-2 \\[2mm]
V^{(a,b,\epsilon)}_{C.S.}=V_{PT}^{(a,b)}(x)+V_{PT}^{(a,b)}(y)+\frac{\epsilon(\epsilon-2)}
{8\sinh^2\frac{x-y}{2}}+\frac{\epsilon(\epsilon-2)}{8\sinh^2\frac{x+y}{2}},\\[2mm]
V_{PT}^{(a,b)}(x)=\frac{(a+b)^2-\frac{1}{4}}{\sinh^2x}-\frac{ab}{\sinh^2\frac{x}{2}}\ .
\end{gather}
Here we have written the Calogero-Sutherland Hamiltonian for arbitrary values of the
coupling $\epsilon = d-2$. It appears when we evaluate the Laplace-Beltrami operator
on the line bundles \eqref{s4ptHA} associated with scalar representations of the
conformal group, see also next section. In the case of $d$-dimensional scalar blocks
there is an additional constant $(d^2-2d+2)/8$ which evaluates to $1/4$ for $d=2$.
According to \cite{Isachenkov:2016gim} the resulting Hamiltonians can be transformed
into the usual Casimir operator \cite{Dolan:2003hv} for scalar 4-point blocks in
2-dimensional conformal field theory, provided the coordinates $x_1 = x$ and
$x_2 = y$ on the Cartan subgroup $A$ are related to the usual variables
$z_1 = z$ and $z_2 = \bar z$ through
\beq \label{zixirel}
z_i= - \sinh^{-2}\frac{x_i}{2}\ .
\eeq
Note that this relation is independent of the dimension $d$.

\section{Example: Seed conformal blocks in 3D}

It has been argued \cite{Costa:2011dw,Iliesiu:2015qra} that all conformal blocks
in 3-dimensional conformal field theory may be obtained from two seed blocks by
application of derivatives. These seed blocks include the usual scalar blocks
along with one type of spinning blocks in which two of the four external fields
transform in a 2-dimensional representation of the rotation group or rather its
universal covering group Spin$(3) = $ SU$(2)$. Our goal is to construct the
Casimir equations for these seed blocks from the Laplace-Beltrami operator on
the 3-dimensional conformal group $\SO(1,4)$. Following the procedure we have
outlined above, we shall end up with two Calogero-Sutherland Hamiltonians. For
scalar blocks, the result agrees with \cite{Isachenkov:2016gim}. In the case of
spinning blocks, on the other hand, we obtain a new formulation of the Casimir
equations that were originally written in \cite{Iliesiu:2015akf}. A verification
that the two sets of Casimir equations are equivalent may be found in Appendix~A.

\subsection{3D scalar blocks}

For scalar blocks the construction of the potential $V$ proceeds exactly as
in our 2-dimensional example in the previous section. In order to build the
Laplacian on the conformal group, we parametrize the two elements $r_i \in
\SO(3)$ in the KAK decomposition \eqref{KAK} through three angles,
\begin{gather}
r_i= \left(
\begin{array}{ccc}
 \cos \phi_i & -\sin \phi_i & 0 \\
 \sin \phi_i & \cos \phi_i & 0  \\
 0 & 0 & 1 \\
\end{array}
\right)
\left(
\begin{array}{ccc}
1 & 0 & 0 \\
0 & \cos \theta_i & -\sin \theta_i \\
0 & \sin \theta_i & \cos \theta_i \\
\end{array}
\right)
\left(
\begin{array}{ccc}
 \cos \psi_i & -\sin \psi_i & 0 \\
 \sin \psi_i & \cos \psi_i & 0  \\
 0 & 0 & 1 \\
\end{array}
\right) \ .
\end{gather}
The angles parametrizing $r_i$ take the values $\phi_{i},\psi_i \in[0,2\pi)$ and
$\theta_i\in[0,\pi]$. The remaining variables $\tau_i$ and $\lambda_i$ run through
the same domain as in our 2-dimensional example.

It is straightforward to compute the metric and to construct the associated
Laplacian. In the case at hand, the Haar measure is given by
\begin{gather}
  d\mu_G= 128(\cosh \tau_1-\cos \tau_2)\sin\theta_1 \sin\theta_2
  \sinh\frac{\tau_1}{2}\sin\frac{\tau_2}{2} \prod_{i=1}^2 d\phi_{i}d\theta_i d\psi_{i}
  d\tau_i d\lambda_i\ .
\end{gather}
If this measure is used to integrate out the angular variables $\phi_i,\psi_i$ and
$\theta_i$, see eq.\ \eqref{Reduction}, and the Laplacian is gauge transformed with
the square root of the function
\begin{equation} \label{m3d}
 m=(\cosh \tau_1-\cos \tau_2)\sinh\frac{\tau_1}{2}\sin\frac{\tau_2}{2}
\end{equation}
as described in eq.\ \eqref{LBAtoHam}, we obtain
\begin{gather}
H=\frac{1}{2}H^{(a,b,1)}_{C.S}+\left.\frac{d^2-2d+2}{8}\right|_{d=3}\ .
\end{gather}
The result is in complete agreement with the Casimir equation for scalar 4-point
functions constructed in \cite{Dolan:2003hv} as was shown in \cite{Isachenkov:2016gim}.

\subsection{3D fermionic seed block}

The fermionic seed block analysed in \cite{Iliesiu:2015akf} involves two spin-$1/2$
fermions at $x_1$ and $x_4$ and two scalar fields that are inserted at $x_2$ and $x_3$.
Consequently, it corresponds to \(\mu_1\otimes\mu'_2=\frac{1}{2}\otimes 0=\frac{1}{2}\)
and $\mu_3\otimes\mu'_4=\frac{1}{2}'$. Explicit parametrisation reads as\footnote{We used that the conjugation of \(r(\theta,\phi,\psi)\in \text{SO}(3)\) with Weyl element \(w=\text{diag}\{1,-1,1,1,-1\}\) acts as \(w r(\theta,\phi,\psi)w=r(-\theta,\phi,\psi)\).}
\begin{gather}
\mathcal{L}(d(\lambda)r) = e^{2a\lambda}\left(
\begin{array}{cc}
 \cos\frac{\theta}{2}e^{i\frac{\phi+\psi}{2}} & i\sin\frac{\theta}{2}e^{i\frac{\phi-\psi}{2}}  \\
 i\sin\frac{\theta}{2}e^{-i\frac{\phi-\psi}{2}} & \cos\frac{\theta}{2}e^{-i\frac{\phi+\psi}{2}}  \\
\end{array} \right), \notag \\
  \mathcal{R}(d(\lambda)r)=e^{-2b\lambda}\left(
\begin{array}{cc}
 \cos\frac{\theta}{2}e^{i\frac{\phi+\psi}{2}} & -i\sin\frac{\theta}{2}e^{i\frac{\phi-\psi}{2}}  \\
 -i\sin\frac{\theta}{2}e^{-i\frac{\phi-\psi}{2}} & \cos\frac{\theta}{2}e^{-i\frac{\phi+\psi}{2}}  \\
\end{array} \right).
\end{gather}
We will continue\footnote{We use the same label \(r\) for an element \(r\in\) SO\((3)\) and its image in SU\((2)\)} to parametrize the left elements $r_l \in$ SU$(2)$ by angles $\phi_l,\psi_l$
and $\theta_l$ and use $\phi_r,\psi_r$ and $\theta_r$ for $r_r \in\,$SU$(2)$. Note that the action
of the right transformations involves $\R(k_r^{-1})$, i.e.\ it contains an additional inversion.
The equivariance law in eq.\ \eqref{Reduction} allows to construct the four components $u_i$ of a
function $f = e^{2a\lambda_l+2b\lambda_r}(u_1,u_2,u_3,u_4)^T$ from a set of functions $u^A_i =
u_i^A(\tau_1,\tau_2)$ on the Cartan subgroup A of the KAK decomposition
\begin{gather}
u_1= e^{\frac{i}{2}(\phi_l-\phi_r-\psi_l-\psi_r)}\left(e^{i\psi_l}\cos\frac{\theta_l}{2}(\cos\frac{\theta_r}{2}\ u_1^A+ie^{i\phi_r}\sin\frac{\theta_r}{2}\ u_2^A)\right.\notag\\
\left.+i\sin\frac{\theta_l}{2}(\cos\frac{\theta_r}{2}\ u_3^A+ie^{i\phi_r}\sin\frac{\theta_r}{2} u_4^A)\right)\\
u_2=e^{\frac{i}{2}(\phi_l-\phi_r-\psi_l+\psi_r)}\left(e^{i\psi_l}\cos\frac{\theta_l}{2}(i\sin\frac{\theta_r}{2}u_1^A+e^{i\phi_r}\cos\frac{\theta_r}{2}u_2^A)+\right. \notag \\ 
\left.+\sin\frac{\theta_l}{2}(-\sin\frac{\theta_r}{2}u_3^A+ie^{i\phi_r}\cos\frac{\theta_r}{2}u_4^A) \right) \\
u_3=e^{-\frac{i}{2}(\phi_l+\phi_r+\psi_l+\psi_r)}\left(ie^{i\psi_l}\sin\frac{\theta_l}{2}(\cos\frac{\theta_r}{2}u_1^A+ie^{i\phi_r}\sin\frac{\theta_r}{2}u_2^A)\right.+\notag\\
\left. +\cos\frac{\theta_l}{2}(\cos\frac{\theta_r}{2}u_3^A+ie^{i\phi_r}\sin\frac{\theta_r}{2}u_4^A) \right)\\
u_4=e^{-\frac{i}{2}(\phi_l+\phi_r+\psi_l-\psi_r)}\left(e^{i\psi_l}\sin\frac{\theta_l}{2}(-\sin\frac{\theta_r}{2}u_1^A+ie^{i\phi_r}\cos\frac{\theta_r}{2}u_2^A)+ \right. \notag \\
\left. + i\cos\frac{\theta_l}{2}(\sin\frac{\theta_r}{2}u_3^A-ie^{i\phi_r}\cos\frac{\theta_r}{2}u_4^A)\right)
\end{gather}
It is now straightforward to work out an expression for the reduction of the
Laplace-Beltrami operator to the Cartan subgroup by inserting the previous list
of formulas for the components of two functions $f$ and $g$ into the general
prescription \eqref{Reduction} and performing the integral over the various
angle variables. After our gauge transformation with the function $m$ given in
eq.\ \eqref{m3d}, the Laplacian takes a block form and nonzero entries can be grouped in two \(2\times 2\) matrices \(H_1=\left(\begin{array}{cc}
H_{22} & H_{23} \\
H_{32} & H_{33} \\
\end{array}\right)\) , \(H_2=\left(\begin{array}{cc}
H_{11} & H_{14} \\
H_{41} & H_{44} \\
\end{array}\right)\)  of Calogero-Sutherland like matrix
Hamiltonians $H_1$ and $H_2$. An additional constant matrix valued gauge
transformation of the form
\begin{gather}\label{LBseedFinal}
\left(\begin{array}{cc}
\frac{1}{\sqrt{2}} & \frac{1}{\sqrt{2}} \\
-\frac{1}{\sqrt{2}} & \frac{1}{\sqrt{2}} \\
\end{array}\right)H_1\left(\begin{array}{cc}
\frac{1}{\sqrt{2}} & -\frac{1}{\sqrt{2}} \\
\frac{1}{\sqrt{2}} & \frac{1}{\sqrt{2}} \\
\end{array}\right)=-\frac{1}{4}\mathcal{M}_1\\
\left(\begin{array}{cc}
\frac{1}{\sqrt{2}} & \frac{1}{\sqrt{2}} \\
-\frac{1}{\sqrt{2}} & \frac{1}{\sqrt{2}} \\
\end{array}\right) H_2\left(\begin{array}{cc}
\frac{1}{\sqrt{2}} & -\frac{1}{\sqrt{2}} \\
\frac{1}{\sqrt{2}} & \frac{1}{\sqrt{2}} \\
\end{array}\right)=-\frac{1}{4}\mathcal{M}_2
\end{gather}
maps these Hamiltonians to the expressions for \(\mathcal{M}_1, \ \mathcal{M}_2\) we
quoted at the end of the Introduction. In the Appendix~A we demonstrate that this
Hamiltonian is equivalent to the Casimir equations derived in \cite{Iliesiu:2015akf}.

\section{Discussion, Outlook and Conclusions}

In this work we build a model of spinning conformal blocks through sections of
a vector bundle over a double-coset of the conformal group to derive Casimir
equations from the Laplace-Beltrami differential operator on $\SO(1,d+1)$. We
argued that the resulting eigenvalue equation takes the form of a
Calogero-Sutherland Schr\"odinger problem for two interacting particles with
spin that move in a 1-dimensional external potential. This potential depends
on the choice of tensor structures and conformal weights of the external
fields and on the dimension $d$ of the space. It was worked out in a few
examples, including the case of 3-dimensional fermionic seed blocks for which
the Casimir equation had originally been derived in \cite{Iliesiu:2015akf}.
The algorithm we described extends to higher dimensions $d\geq 4$ with only
one significant change, namely that the KAK decomposition is no longer
unique. In order to fix the issue, one can restrict one of the factors $K$
to the coset space $K/B$ where $B = SO(d-2)$. At the same time, the fibers
of the relevant vector bundles must be projected to the subspace of
$\SO(d-2)$ invariants. We will describe this in more detail in a forthcoming
paper \cite{SS} on Casimir equations for 4-dimensional seed blocks, see
\cite{Echeverri:2016dun}.%
\medskip

There are a number of other extensions that seem worth pursuing. To begin
with, it would be interesting to work out the Calogero-Sutherland Hamiltonians
for blocks of scalar and tensor fields in supersymmetric theories. Most
of the existing work on Casimir equations in such theories focuses on
correlation functions of BPS operators. If all four external operators
are BPS, the Casimir equations resemble those for scalar blocks in bosonic
theories \cite{Fitzpatrick:2014oza,Bobev:2015jxa,Lemos:2015awa} and hence
they can be cast into a Calogero-Sutherland like form. Things become more
interesting when we admit non-BPS operators. There are only a few cases in
which the Casimir equations for such setups have been worked out, see e.g.\
\cite{Fitzpatrick:2014oza} and \cite{Cornagliotto} for 2-dimensional theories
with $\mathcal{N}=1$ and $\mathcal{N}=2$ supersymmetry, respectively.

Other interesting extensions concern correlation functions of local operators
in the presence of boundaries and defects. All these scenarios can be cast into
the framework we outlined above. The main difference is that the left and right
subgroups $K_l=K$ and $K_r=K$ that we divided by above must be chosen according
to the geometry of the configuration. In particular, they are usually not equal
to each other any longer. If we want to describe conformal blocks for two bulk
fields in the presence of a boundary, for example, we have to consider the coset
$K_l\setminus G/K$ where $K_l = \SO(1,d)$ is the $d-1$ dimensional conformal
group and $K_r = K$ is the same as before. We plan to work out a number of such
examples and to compare with known Casimir equations whenever they are available,
see e.g. \cite{Liendo:2012hy,Billo:2016cpy,Gadde:2016fbj}.

For technical reasons we worked with the principle series representations of
conformal weight $\Delta=d/2+ic$ and performed an analytic continuation to the
real weights of local fields only in the very last step. On the other hand,
there could exist direct applications to a broader class of operators. In \cite{Balitsky:2013npa} one of authors introduced a new class of nonlocal
light-ray operators that realize the principle series representation of
sl$(2|4)$ and then calculated their correlation function in BFKL regime 
\cite{Balitsky:2015tca,Balitsky:2015oux}. It would be very tempting
to extend the bootstrap programme to such type of operators.%  
\medskip

What we have explored here so far is a very universal new approach to
conformal blocks that may be applied to a wide variety of setups, including
boundaries, defects and supersymmetric theories. As we have also seen
in the example of the 3D seed blocks, it casts the Casimir equations
into a new and often simpler looking form. But the main interest of our
approach is that it embeds the theory of conformal blocks into the rich
world of (super-~)integrable quantum systems. In the case that is relevant
for conformal blocks of scalar fields, super-integrability is firmly
established, see \cite{Isachenkov:2016gim} and references therein, though
it still remains to be exploited \cite{IsachenkovI,IsachenkovII}. The
analysis presented above suggests that the connection between blocks and
integrability goes much deeper and, in particular, also includes blocks
with external tensor fields. Let us explain this in a bit more detail.
Harmonic analysis on a Lie group is usually not an integrable problem.
In fact, the number of independent commuting (differential) operators
is given by the rank $r$ of the group and hence is much smaller than the
number \textit{dim}$G$ of coordinates. In performing the reduction to coset
geometries, however, we reduce the number of coordinates while keeping the
same number of commuting operators unless they start to become dependent.
The conformal group possesses $r=[d+2/2]$ independent Casimir elements. So,
when we reduce to our double coset, these outnumber the coordinates and
hence the quantum mechanical system becomes integrable at least before we
add spin degrees of freedom. The first case in which there are infinitely
many spinning conformal seed blocks appears in $d=4$ dimensions. At this
dimension, the number $r$ of Casimir invariants jumps from $r=2$ for $d<4$
to $r=3$, i.e. there is one more Casimir invariant than there are cross
ratios or coordinates on the double coset. It seems likely that the
additional Casimir invariant makes the corresponding spinning quantum
mechanical systems integrable. For the spinning \textit{A}$_n$
Calogero-Sutherland Hamiltonians which are associated to bundles over
adjoint coset spaces $G/G$, super-integrability (or degenerate integrability) has
recently been proven in \cite{Reshetikhin:2015rba}. It remains to extend
such an analysis to \textit{BC}$_n$ root systems and thereby to spinning
conformal blocks.

Super-integrability is a powerful feature. As is well known from the
Runge-Lenz vector of the hydrogen atom, the spectrum generating symmetries
of super-integrable systems can make them algebraically solvable. In the
case of conformal blocks, all the known recurrence relations \cite{Dolan:2011dv}
are direct consequences of super-integrability \cite{IsachenkovI,IsachenkovII}.
We believe that the remarkable formulas for 4-dimensional seed blocks
that were found in \cite{Echeverri:2016dun} can be understood through the
super-integrability of the associated Calogero-Sutherland systems. If this
was true, it would pave the way for extensions, e.g. to other dimensions.
We plan to return to these questions in future research.

\begin{acknowledgments}
\label{sec:acknowledgments}
We thank Lazlo Feher, Denis Karateev, Zohar Komargodski, Madalena Lemos, Pedro Liendo, Vincent Pasquier, Joao Penedones and Didina Serban for comments and discussions. E.Sobko thanks the Department of Mathematics, King's college London, and in particular the DESY Theory Group for their hospitality during the earlier stages of this research. The work
of E. Sobko was supported by the grant "Exact Results in Gauge and String Theories"
from the Knut and Alice Wallenberg
foundation and the People Programme (Marie Curie Actions) of the European Union's Seventh
Framework Programme FP7/2007-2013/ under REA Grant Agreement No 317089 (GATIS). The work
of M. Isachenkov was supported by an Israel Science Foundation center for excellence
grant, by the I-CORE program of the Planning and Budgeting Committee and the Israel
Science Foundation (grant number 1937/12), by the Minerva foundation with funding from
the Federal German Ministry for Education and Research, by a Henri Gutwirth award from
the Henri Gutwirth Fund for the Promotion of Research, by the ISF within the ISF-UGC
joint research program framework (grant no. 1200/14) and by the ERC STG grant 335182.
\end{acknowledgments}

\appendix

\section{Comparing with 3D fermionic blocks from \cite{Iliesiu:2015akf}}\label{ApConfStrOf3LR}
In this section we rewrite Casimir equations for the fermionic seed blocks that were derived
in \cite{Iliesiu:2015akf} as a matrix valued Calogero-Sutherland like eigenvalue equation that
may be compared with the expressions we obtained by our reduction from the Laplacian on the
conformal group. We start by reproducing the equations (A.10) from \cite{Iliesiu:2015akf}
\begin{gather}
\left[ \left(
\begin{array}{cc}
 \mathcal{L}^+_D & \mathcal{L}^+_A \\
 \mathcal{L}^+_A & \mathcal{L}^+_D \\
\end{array}
\right)+
 \left(
\begin{array}{cc}
 0 & -\frac{8r(a+b)}{1+r^2-2r\eta} \\
 0 & -\frac{8r(\eta-2r+r^2\eta)(a+b)}{(1+r^2-2r\eta)2} \\
\end{array}
\right)\right]
\left(
\begin{array}{cc}
\tilde{g}^1_{\Delta,l}  \\
\tilde{g}^2_{\Delta,l}  \\
\end{array}
\right)
=C_{\Delta,l}
\left(
\begin{array}{cc}
\tilde{g}^1_{\Delta,l}  \\
\tilde{g}^2_{\Delta,l}  \\
\end{array}
\right)\label{A1stSyst}\\
\left[ \left(
\begin{array}{cc}
 \mathcal{L}^-_D & \mathcal{L}^-_A \\
 \mathcal{L}^-_A & \mathcal{L}^-_D \\
\end{array}
\right)+
 \left(
\begin{array}{cc}
 -\frac{8r(\eta+2r+r^2\eta)b}{(1+r^2+2r\eta)^2} & \frac{8ra}{1+r^2+2r\eta} \\
 \frac{8rb}{1+r^2+2r\eta} & -\frac{8r(\eta+2r+r^2\eta)a}{(1+r^2+2r\eta)^2} \\
\end{array}
\right)\right]
\left(
\begin{array}{cc}
\tilde{g}^3_{\Delta,l}  \\
\tilde{g}^4_{\Delta,l}  \\
\end{array}
\right)
=C_{\Delta,l}
\left(
\begin{array}{cc}
\tilde{g}^3_{\Delta,l}  \\
\tilde{g}^4_{\Delta,l}  \\
\end{array}
\right)\label{A2ndSyst}
\end{gather}
where
\begin{gather}
\mathcal{L}_D^\pm=r^2\partial^2_r+(\eta^2-1)\partial^2_\eta+\notag\\
+\left(\frac{-8r^2\eta(1-r^2)(a+b)}{(1+r^2-2r\eta)(1+r^2+2r\eta)}-
\frac{r(1+3r^2)}{1-r^2}-\frac{r(1-r^2)(1+r^2\mp2r\eta)}{(1+r^2+2r\eta)
  (1+r^2-2r\eta)} \right)\partial_r\notag\\
+\left(\frac{-8(\eta^2-1)(r^3+r)(a+b)}{(1+r^2+2\eta r)(1+r^2-2\eta r)}+
\frac{(3\eta(1+r^2)\pm 2r(4\eta^2-1))(1+r^2\mp2r\eta)}{(1+r^2+2\eta r)
  (1+r^2-2\eta r)} \right)\partial_\eta\notag\\
+\left(\frac{3}{4}-\frac{16abr(\eta+2r+r^2\eta)}{(1+r^2+2r\eta)^2}\right)\\
\mathcal{L}^\pm_A=\frac{2r^2}{1-r^2}\partial_r\pm\partial_\eta
\end{gather}
and \(\Delta_{12}=-2a\), \(\Delta_{43}=-2b\). To begin rewriting these
expressions, we perform the following change of variables
\begin{gather}\label{ChangeOfVar}
\begin{align}
&r=e^{\frac{x+y}{2}},\\
&\eta=-\cosh\frac{x-y}{2}
\end{align}
\end{gather}
After this change of variables the system of equations (\ref{A1stSyst}-\ref{A2ndSyst})
continues to possess the matrix form
\begin{gather}
\tilde{\mathcal{M}}_{1}\left(
\begin{array}{cc}
\tilde{g}^1_{\Delta,l}  \\
\tilde{g}^2_{\Delta,l}  \\
\end{array}
\right)
=C_{\Delta,l}
\left(
\begin{array}{cc}
\tilde{g}^1_{\Delta,l}  \\
\tilde{g}^2_{\Delta,l}  \\
\end{array}
\right), \notag \\
\tilde{\mathcal{M}}_{2}\left(
\begin{array}{cc}
\tilde{g}^3_{\Delta,l}  \\
\tilde{g}^4_{\Delta,l}  \\
\end{array}
\right)
=C_{\Delta,l}
\left(
\begin{array}{cc}
\tilde{g}^3_{\Delta,l}  \\/
\tilde{g}^4_{\Delta,l}  \\
\end{array}
\right)
\end{gather}
Explicit formulas for the matrices $\tilde{\mathcal{M}}_i$ of differential operators in $x$ and
$y$ are easily worked out. Once they are derived, we perform the following transformations
\begin{gather}
\mathcal{M}_{1}=-\frac{1}{2}\begin{pmatrix}
\chi_1(x,y) & \chi_2(x,y) \\
-\chi_1(x,y) & \chi_2(x,y)
\end{pmatrix}^{-1}\tilde{\mathcal{M}}_{1}\begin{pmatrix}
\chi_1(x,y) & \chi_2(x,y) \\
-\chi_1(x,y) & \chi_2(x,y)
\end{pmatrix}\\
\mathcal{M}_{2}=-\frac{1}{2}\begin{pmatrix}
\chi_3(x,y) & \chi_4(x,y) \\
-\chi_3(x,y) & \chi_4(x,y)
\end{pmatrix}^{-1}\tilde{\mathcal{M}}_{2}\begin{pmatrix}
\chi_3(x,y) & \chi_4(x,y) \\
-\chi_3(x,y)& \chi_4(x,y)
\end{pmatrix}
\end{gather}
where
\begin{gather}
  \chi_1(x,y)=\frac{{\cosh \frac{x}{2}}^{-a-b}\sinh\frac{x}{2}^{-\frac{1}{2}+a+b}
    \cosh\frac{y}{2}^{-a-b}\sinh\frac{y}{2}^{-\frac{1}{2}+a+b}}{(\cosh\frac{y}{2}
    -\cosh\frac{x}{2})^{\frac{3}{2}}(\cosh\frac{y}{2}+\cosh\frac{x}{2})^{\frac{1}{2}}},
  \notag\\
  \chi_2(x,y)=\frac{\cosh\frac{x}{2}^{-a-b}\sinh\frac{x}{2}^{-\frac{1}{2}+a+b}
    \cosh\frac{y}{2}^{-a-b}\sinh\frac{y}{2}^{-\frac{1}{2}+a+b}}{(\cosh\frac{y}{2}
    -\cosh\frac{x}{2})^{\frac{1}{2}}(\cosh\frac{y}{2}+\cosh\frac{x}{2})^{\frac{3}{2}}},
  \notag\\
  \chi_3(x,y)=\frac{\cosh\frac{x}{2}^{-\frac{1}{2}-a-b}\sinh\frac{x}{2}^{a+b}
    \cosh\frac{y}{2}^{-\frac{1}{2}-a-b}\sinh\frac{y}{2}^{a+b}}{(\sinh\frac{x}{2}-
    \sinh\frac{y}{2})^{\frac{1}{2}}(\sinh\frac{x}{2}+\sinh\frac{y}{2})^{\frac{3}{2}}},
  \notag\\
  \chi_4(x,y)=\frac{\cosh\frac{x}{2}^{-\frac{1}{2}-a-b}\sinh\frac{x}{2}^{a+b}
    \cosh\frac{y}{2}^{-\frac{1}{2}-a-b}\sinh\frac{y}{2}^{a+b}}{(\sinh\frac{x}{2}
    -\sinh\frac{y}{2})^{\frac{3}{2}}(\sinh\frac{x}{2}+\sinh\frac{y}{2})^{\frac{1}{2}}}.
\end{gather}
After this transformation, the system \ref{A1stSyst}-\ref{A2ndSyst} now reads
\begin{gather}
\mathcal{M}_{1}\left(
\begin{array}{cc}
g^1_{\Delta,l}  \\
g^2_{\Delta,l}  \\
\end{array}
\right)
=-\frac{C_{\Delta,l}}{2}
\left(
\begin{array}{cc}
g^1_{\Delta,l}  \\
g^2_{\Delta,l}  \\
\end{array}
\right), \notag \\
\mathcal{M}_{2}\left(
\begin{array}{cc}
g^3_{\Delta,l}  \\
g^4_{\Delta,l}  \\
\end{array}
\right)
=-\frac{C_{\Delta,l}}{2}
\left(
\begin{array}{cc}
g^3_{\Delta,l}  \\
g^4_{\Delta,l}  \\
\end{array}
\right)
\end{gather}
where the operators \(\mathcal{M}_{1},\ \mathcal{M}_{2}\) are the same as in the
Introduction and the new functions \(\{g^i_{\Delta,l}\}\) are related to conformal
blocks \(\{\tilde{g}^i_{\Delta,l}\}\) as
\begin{gather}
\left(
\begin{array}{cc}
g^1_{\Delta,l}  \\
g^2_{\Delta,l}  \\
\end{array}
\right)=\begin{pmatrix}
\chi_1(x,y) & \chi_2(x,y) \\
-\chi_1(x,y) & \chi_2(x,y)
\end{pmatrix}^{-1}
\left(
\begin{array}{cc}
\tilde{g}^1_{\Delta,l}  \\
\tilde{g}^2_{\Delta,l}  \\
\end{array}
\right)  \\
\left(
\begin{array}{cc}
g^3_{\Delta,l}  \\
g^4_{\Delta,l}  \\
\end{array}
\right)=\begin{pmatrix}
\chi_3(x,y) & \chi_4(x,y) \\
-\chi_3(x,y) & \chi_4(x,y)
\end{pmatrix}^{-1}
\left(
\begin{array}{cc}
\tilde{g}^3_{\Delta,l}  \\
\tilde{g}^4_{\Delta,l}  \\
\end{array}
\right)
\end{gather}

\bibliographystyle{JHEP.bst}
%\bibliography{mybib_VK}

%\begin{thebibliography}{99}
\bibliographystyle{plain}
\bibliography{literatureHA}
%\end{thebibliography}

\printindex

\end{document}